\documentclass[journal,onecolumn]{IEEEtran}
\usepackage{epsfig}
\usepackage{graphicx}
\usepackage{graphics}
\usepackage{epstopdf} 
\usepackage{tikz}
\usetikzlibrary{calc,shapes.geometric}
\usetikzlibrary{arrows,snakes,backgrounds}
\usepackage{pgf, tikz}
\usetikzlibrary{arrows, automata, positioning}
\usetikzlibrary{quotes,angles}
\usepackage{amssymb}

\usepackage{tcolorbox}
\tcbset{colframe=black,colback=white,colupper=black,
fonttitle=\bfseries,nobeforeafter,center title,size=small,arc=0pt,outer arc=0pt}

\usepackage{epsfig}
\usepackage{graphicx}
\usepackage{graphics}
\usepackage{epstopdf} 
\usepackage{tikz}
\usetikzlibrary{calc,shapes.geometric}
\usetikzlibrary{arrows,snakes,backgrounds}
\usepackage{pgf, tikz}
\usetikzlibrary{arrows, automata, positioning}
\usetikzlibrary{quotes,angles}
\usepackage{amssymb}
\usepackage{amsmath}
\usepackage{xcolor}
\usepackage{amssymb}
\usepackage{multirow}
\usepackage{subfigure}
\usepackage{cite}

\usepackage{tcolorbox}
\tcbset{colframe=black,colback=white,colupper=black,
fonttitle=\bfseries,nobeforeafter,center title,size=small,arc=0pt,outer arc=0pt}

\def\nexto{\kern -0.54em}

\def\prob{{\rm {I\ \nexto P}}}

\begin{document}

\title{Selecting the Number of Sensor Scans in Surveillance Operations}
\author{Graham  V. Weinberg\\
Graham.Weinberg@dst.defence.gov.au
 }
\maketitle


\begin{abstract}
Searching for concealed threats within a surveillance region is an important role for military sensors. One prominent case is the search for a submarine by a helicopter deploying a dipping sonar in anti-submarine warfare. Another is the utilisation of an uncrewed aerial vehicle for remote detection of land mines.  These platforms will deploy an appropriate sensor to search within a given region for a potential threat. The sensors employed may not provide continuous surveillance, but may have active periods where the inherent sensor is scanning for threats. The reasons for disjoint activation periods may be both functional as well as tactical.
Therefore, given a specified search pattern under which the surveillance platform operates, the question of the number of sensor scans required becomes relevant for search mission pre-planning. Hence it is not only important to develop a modelling framework to assess performance but it is of relevance to consider whether a minimum number of sensor scans can be determined to guarantee a desired level of performance. Consequently this paper introduces a framework for this and illustrates how it can be utilised in practice.
\end{abstract}


\setlength {\abovedisplayskip} {6pt plus 3.0pt minus 4.0pt}
\setlength {\belowdisplayskip} {6pt plus 3.0pt minus 4.0pt}

\section{Introduction}
Successful detection of low-signature threats from remote surveillance platforms is an important capability for the modern fighting force. This is particularly problematic since there is a juxtaposition between improvements in both stealth and sensing technology. Nonetheless there is a need to provide a performance prediction framework to assess the success in detecting relevant threats with sensor systems. The case of detection of a submarine from a helicopter utilising a dipping sonar is a good example of the complexities in successful detection \cite{danskin68, yoash18, young22}. Dipping sonars are omnidirectional transceivers which are deployed for a short period of time in anti-submarine warfare (ASW) operations. The sea environment has a serious effect on the relative success of detection, since sonar pings are attenuated by temperature, sea-state, salinity and bathymetry \cite{urick}. In addition to this, the sound speed profile varies with range and so while a submarine is already a low acoustic-signature threat, the environment itself may impede successful detection \cite{urick}.

In a similar vein, an uncrewed aerial vehicle (UAV) may be equipped with land mine detection functionality, and so is capable of remote detection of land-mines \cite{sabatier06, kovacs22}. In the case of acoustic sensors the environment will also have a negative impact on successful detection but not as severe as in the undersea environment. When utilising optical or radar sensors, environmental characteristics will also contribute to detection performance degradation. 

Hence the approach to be adopted in this study is to assume that in a given operational environment there is a known detection range, such that if a threat falls within this range it will be sensed by the platform. This is known as adopting a ``cookie-cutter'' model for target detection \cite{craparo}. Doing this allows approximate rules of thumb to be established to assess operational capability. For a given operational environment, and with a platform employing a specific sensor, one may apply well-known engineering functions to determine maximal detection ranges. In the case of radar, this is the radar-range equation, while for acoustic sensors it is the well-known sonar equation \cite{urick, stimpson}. Based upon sensor attributes and environmental factors one may then extract the detection range and use this as a basis for performance prediction and assessment. 

The objective of this paper is to characterise the threat detection problem in terms of sensor scans and utilising a maximal detection range assumption. In particular, supposing that a surveillance platform is deploying a sensor, which is switched on for fixed periods to scan for threats, a model is developed to determine the probability of detection.  The principle of fixed time scans has its origins in the ASW aspect of this study. However, if a UAV is deploying a radar then it can become the target of a radio frequency seeker and so short periodic scans may sustain the platform's survivability. Once an expression for the probability of detection is produced, one may proceed to develop a rule of thumb to determine the minimum number of sensor scans required to guarantee a desired level of performance. 

It is important to note that this study does not consider the problem from the perspective of determining an optimal sensor platform trajectory but instead is concerned with performance assessment with a given trajectory. From an ASW perspective other studies have focused on dipping pattern analysis \cite{yoash18} and search region coverage \cite{craparo}.
This study's contribution is to show how a simple mathematical model may be developed such that it can be used to provide insights on sensor performance in a number of relevant applications.

The developments in this study will then be illustrated with a simple example from ASW operations, to demonstrate how it can be applied to a specific sensor on a given platform.

\section{Surveillance Performance Model}
It is assumed that a threat is located somewhere within a bounded surveillance region, and that this region is viewed as a two-dimensional plane for modelling simplicity.
Suppose that the surveillance path within this region, traversed by the platform deploying the sensor, is parameteristed through $\{(x(t), y(t)), t \in {\cal T}\}$ which represents a curve in two-dimensional space, and that the sensor is deployed at times $\tau_1, \tau_2, \ldots, \tau_N$, with ${\cal T}$ an interval of time. Hence the total number of surveillance scans is $N$. For brevity define $x_j = x(\tau_j)$ and $y_j = y(\tau_j)$ for $j \in \{1, 2, \ldots, N\}$.

Suppose that at scan time $\tau_j$ the threat will be detected by the sensor if it lies within a maximal detection range, given by $\epsilon > 0$. 
Then if the threat is located at coordinates $(x_0, y_0)$  it will be detected by the sensor if
\begin{equation}
d_j = d( (x_j, y_j), (x_0, y_0)) = \sqrt{ (x_j - x_0)^2 + (y_j - y_0)^2} < \epsilon. \label{eqA}
\end{equation}
Expression \eqref{eqA} is the Euclidean distance between the sensor and threat location.
The threat location $(x_0, y_0)$ may be modelled through a statistical distribution, such as assuming that it is uniform in both coordinates. This will be clarified in the analysis to follow.

Given that there are $N$ scans the probability that the threat is not detected is exactly the probability that $d_j \geq \epsilon$ for each $j \in \{1, 2, \ldots, N\}$. If it is assumed that each scan is independent then 
\begin{equation}
\prob({\mbox{Threat not detected}}) = \prod_{j=1}^N \prob(d_j \geq \epsilon).
\end{equation}
Hence by complementation it follows that 
\begin{equation}
\prob({\mbox{Threat detected}}) = 1 - \prod_{j=1}^N \prob(d_j \geq \epsilon).
\label{eqB}
\end{equation}
Under the assumption that the sensor deployment times are deterministic it is not difficult to express the probability in \eqref{eqB} in terms of the conditional distribution of the threat location. However, \eqref{eqB} may also be evaluated through Monte Carlo methods in specific examples.

\begin{figure}[h]
\includegraphics[width=10.5 cm]{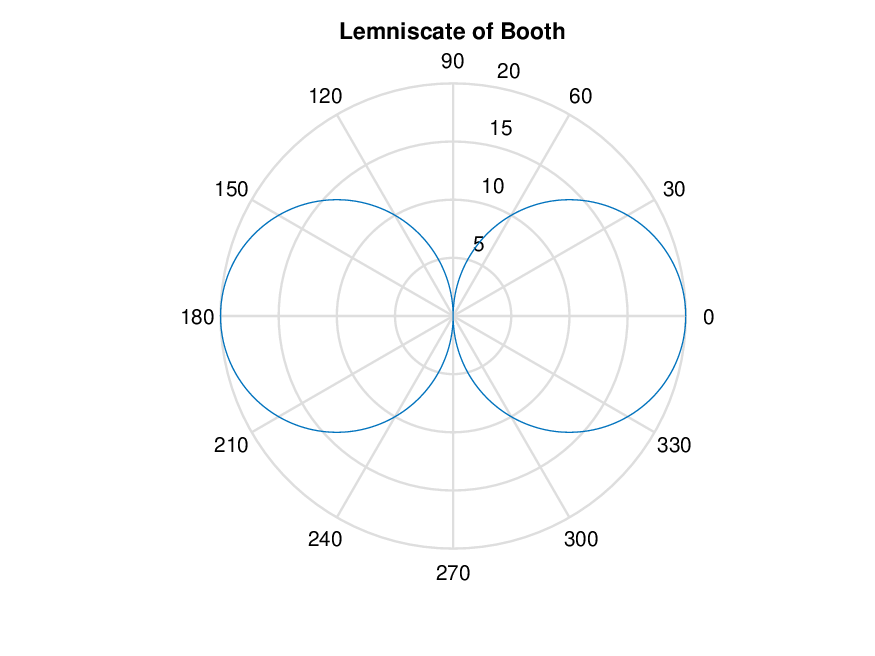}
\caption{Leminscate surveillance path polar plot.\label{fig1}}
\end{figure}

\begin{figure}[h]
\includegraphics[width=10.5 cm]{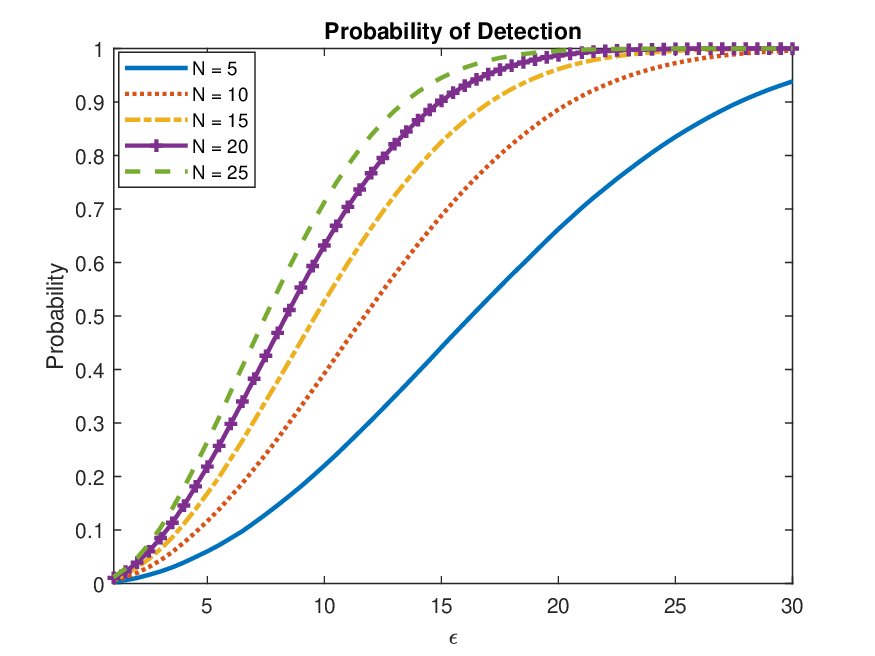}
\caption{Detection performance as a function sensor effective range $\epsilon$ for various number of scans $N$.\label{fig2}}
\end{figure}

It is now worth providing a specific example of \eqref{eqB}. Towards this end, suppose that the surveillance region is a square defined through the Cartesian product $[-40, 40] \times [-40, 40]$ and that the threat's location is assumed to be uniformly distributed in this region. The units here are arbitrary but can be specialised in terms of a specific application.  The surveillance path is taken to be a lemniscate of Booth given by the polar equation $r = 20 \cos(\theta)$, where the angle $\theta$ corresponds to uniformly spaced scan points. Hence if there are $N$ surveillance scans then $\theta_j = \frac{2\pi}{N}j$ for $j \in \{1, 2, \ldots, N\}$. Figure \ref{fig1} provides a polar plot of this path. The selection of this path has been motivated by the fact that it has the potential to provide better regional coverage than a circular path.  Figure \ref{fig2} plots the probability of detection as a function of $\epsilon$ for the latter ranging up to 30. Expression \eqref{eqB} has been evaluated using Monte Carlo simulation with $10^5$ runs for each point probability. The figure shows the probability of detecting the threat, as a function of the number of scans and the maximal detection range $\epsilon$. As can be observed, increasing the number of scans will improve performance as expected,  and as the sensor's maximal detection range increases there is also a corresponding improvement in performance.

For an explicit example, if the sensor has a range of 10 units then at least 15 scans is required to guarantee a probability of detection of at least 50\%. More than 25 scans would be required to guarantee a probability of detection of at least 70\%.  If the sensor had a range of 20 units then only 5 sensor scans would be required to achieve a probability of detection of approximately 60\%.  Additionally, 15 scans would be sufficient to ensure that this probability increases to at least 90\%.

\begin{figure}[h]
\includegraphics[width=10.5 cm]{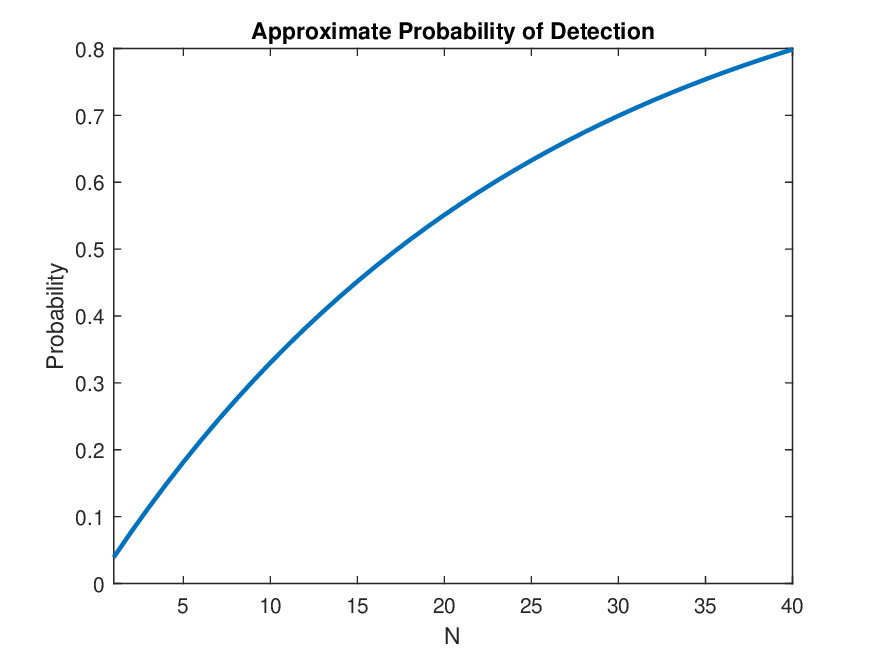}
\caption{Approximate detection performance, based upon \eqref{eqC}, where $N$ is the number of sensor scans. \label{fig3}}
\end{figure}

Although plots such as that in Figure \ref{fig2} may be used to assess performance of particular systems it is worth considering whether an approximate guide to the number of surveillance scans can be produced. In effect one would like to detemine a suitable minimum $N$ so that \eqref{eqB} is at a desired level.
In order to derive an appropriate rule of thumb for the number of required sensor scans, suppose that the surveillance region is a square in two-dimensional space given by the Cartesian product $[-\delta, \delta] \times [-\delta, \delta]$ for some known $\delta>0$. Assume that the surveillance path is contained within this region such that for any scan point $(x_j, y_j)$ the circle with radius $\epsilon$ and centre at this point lies within the surveillance region boundary. Then the probability that the threat is detected at the $j$th scan is the probability that it lies within the circle, centre $(x_j, y_j)$ with radius $\epsilon$. From a geometric perspective this is exactly the area of the circle to that of the surveillance region. Therefore 
\begin{equation}
\prob((x_0, y_0) \mbox{ lies in the circle}) = \frac{\pi \epsilon^2}{4 \delta^2}.
\label{eqD}
\end{equation}
This is independent of the scan position provided the underlying circle lies within the surveillance region. The complement of expression \eqref{eqD} may now be applied to \eqref{eqB}
and consequently it follows that 
\begin{equation}
\prob({\mbox{Threat detected}}) = 1  - \left( 1 - \frac{\pi \epsilon^2}{4 \delta^2}\right)^N.
\label{eqC}
\end{equation}
The validity of \eqref{eqC} is dependent on the assumption that the series of circles do not overlap. In order to ensure this, is is necessary to suppose that the ratio of $\epsilon$ to $\delta$ is sufficiently small. As an example, suppose that the square of this ratio is 0.05, then Figure \ref{fig3} plots \eqref{eqC} as a function of $N$. From this figure one may determine roughly the number of sensor scans for a desired probability of detection, regardless of the path over which the surveillance platform operates. However, results derived from this curve will be somewhat conservative. As an example, the figure suggests that for a probability of detection of at least 0.7 it is necessary to have at least 30 sensor scans.

\section{Application}
In this final section an application of the results of the preceding section is considered. In particular, the problem of ASW with a dipping sonar is examined, and the following is based upon results discussed in \cite{tyler92}. For simplicity, suppose that a helicopter is deploying an omnidirectional dipping sonar, operating in a narrowband mode within a search region which is a square of length 400 nautical miles (nmi). The location of the submarine is distributed uniformly within this square. As in the previous analysis, assume that the helicopter is travelling on a lemniscate path with polar equation $r = 16\cos(\theta)$. The equation's parameter 16 has been selected based upon the operational range of the MH-60R Seahawk multimission naval helicopter whose maximum range  is approximately 450 nmi \cite{romeo}. Using calculus it can be shown that the path length of the given lemniscate is $\pi/2 \times 16^2 \approx 402$ nmi so that the helicopter should be able to traverse the full lemniscate path in one mission.

From the active sonar equation, the submarine will be detected if the signal excess (SE) exceeds zero. Under the assumption that the transmission loss (TL) from source to target is the same from target to source, the sonar equation is 
\begin{equation}
SE = SL-2TL+TS-RL-DT, \label{eqD2}
\end{equation}
where SL is the source level, TS is the target strength, RL is the reverberation loss and DT is the detection threshold \cite{ainslie}. 
This equation is constructed from the logarithm of the signal to noise ratio, and all its parameters are in dB. Suppose that the environment has a constant sound speed profile ($c = 3355$ miles per hour) and that RL is a constant multiple of the SL. The simplified model for TL is given by
\begin{equation}
TL = 66 + 10 \log_{10}(R) + 2\alpha R, \label{eqE}
\end{equation}
where $R$ is the distance to the target and the absorption coefficient $\alpha$ is given by
\begin{equation}
\alpha = 0.1 f^2/(1+f^2) + 40f^2/(4100+f^2) + 2.75\times 10^{-4} f^2 + 0.003,
\label{eqF}
\end{equation}
where $f$ is the ping frequency.

The model for TS, based upon a cylindrical approximation for the shape of a submarine,  is given by
\begin{equation}
TS = (a L^2/2\lambda)(\sin(\beta)/\beta)^2 \cos(\psi)^2, \label{eqG}
\end{equation}
where $a$ is the radius of a cylinder of length $L$, $\lambda$ is the wavelength, $\psi$ is the aspect angle off-beam and $\beta = 2\pi/\lambda L \sin(\psi)$.

Finally the DT is given by
\begin{equation}
DT = 10 \log_{10}(d/2T), \label{eqH}
\end{equation}
where $d$ is the detection index and $T$ is the duration of the waveform. 

\begin{figure}[h]
\includegraphics[width=10.5 cm]{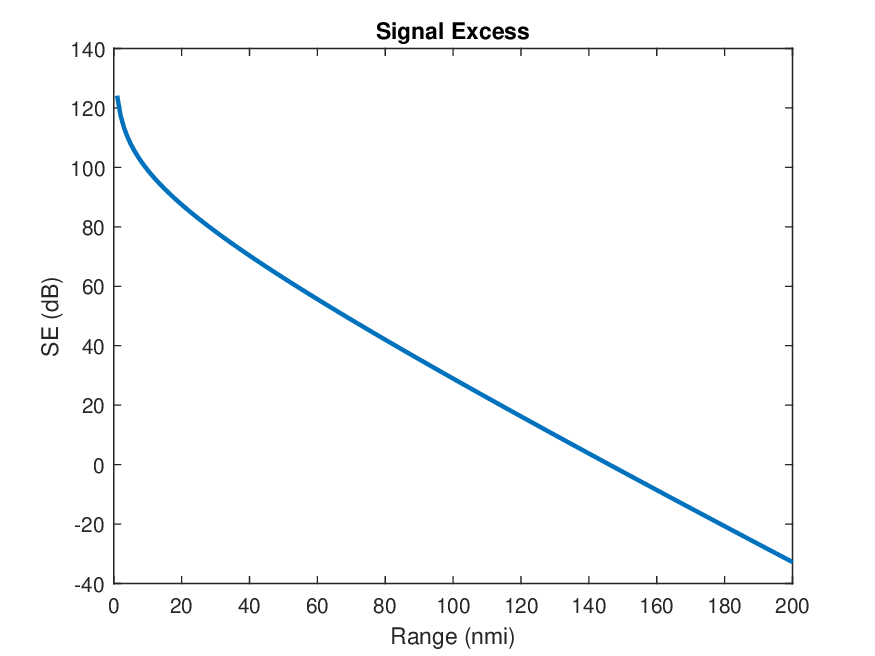}
\caption{A plot of the signal excess (SE) as a function of range, for the model parameterisations utilised in the study. Detection will occur in cases where the SE exceeds zero.\label{fig4}}
\end{figure}

\begin{figure}[h]
\includegraphics[width=10.5 cm]{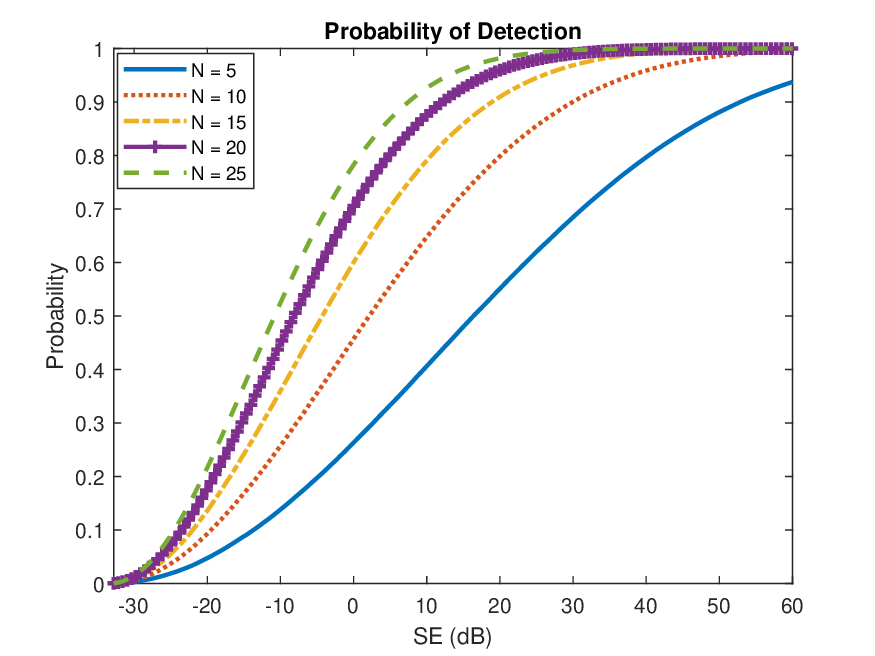}
\caption{Probability of detection as a function of SE for the parameterisations employed in the previous figure. A selection of number of sensor scans $N$ is shown to illustrate the variation of performance with this parameter.\label{fig5}}
\end{figure}

Expressions \eqref{eqE} - \eqref{eqH} may be applied to \eqref{eqD2} to produce range-dependent SE values.
In the following $f = 10$ Hz with $d = 25$ and $T = 100$ s. This choice for $d$ ensures that the corresponding false alarm probability is $10^{-4}$.
For the cylindrical model of the threat, $L =300$ ft, $a = 15$ ft and $\psi = \pi/4$.
The SL is taken to be 250 dB and the RL is one tenth of this. This selection of SL has been partially motivated by discussions in \cite{donaldson17} for the MH-60R helicopter, while the RL has been an arbitrary selection. Figure \ref{fig4} plots \eqref{eqD2} as a function of the range to the threat.
This is then adapted to produce Figure \ref{fig5}, which graphs \eqref{eqB} for five cases of $N$ as considered previously. 
The figure shows that for small SE more scans are required to ensure adequate probability of detection, while for threats with a stronger SE one may utilise less scans to achieve likely detection. As tangible examples from Figure \ref{fig4} at a range of 140 nmi the SE is approximately 0 dB. Based upon Figure \ref{fig5} it is clear that $N = 20$ scans is sufficient to guarantee a probability of detection of at least 70\%, while to increase this to at least 80\% one must increase the number of scans to at least 25. By contrast, at a range of 50 nmi the SE is approximately 60 dB and detection is guaranteed (with a probability of approximately 90\%) with only $N = 5$ scans. 

\section{Conclusions and Further Work}
The purpose of this study was to introduce a modelling framework within which one may assess the probability of detection of a threat using a sensor system which scans at given times. 
The analysis provided a graphical means of assessing performance as a function of number of scans and maximal detection range. An approximate rule of thumb provided a conservative estimate of the number of scans independent of the surveillance platform's trajectory. Finally a specific ASW model was used to illustrate how the techniques could be combined with more detailed expressions for maximal detection ranges.

Further work could focus on producing a more accurate rule of thumb for approximate performance, as well as the application of the results to more comprehensive models in sonar system performance. In addition to this one could explore the problem of UAV-based detection of land mines.

\end{document}